\documentclass[twocolumn,showpacs,preprintnumbers,amsmath,amssymb,pra,superscriptaddress]{revtex4}

\usepackage{graphicx}% Include figure files
\usepackage{epsfig}
\usepackage{dcolumn}% Align table columns on decimal point
\usepackage{textcomp}
\usepackage{bm}% bold math

\def\rm#1{\mathrm{#1}}

\begin{document}

%\preprint{APS/123-QED}

%\preprint{Draft in Confidence}
\title{Identifying a Two-State Hamiltonian in the Presence of Decoherence} 

\author{Jared~H.~Cole}
 \email{j.cole@physics.unimelb.edu.au}
\author{Andrew D. Greentree}
\affiliation{%
Centre for Quantum Computer Technology, School of Physics, The University of Melbourne, Melbourne, Victoria 3010, Australia.
}%
\author{Daniel~K.~L.~Oi}
\author{Sonia~G.~Schirmer}
\affiliation{%
Department of Applied Mathematics and Theoretical Physics, The University of Cambridge, Wilberforce Road, Cambridge CB3 0WA, United Kingdom
}%

\author{Cameron~J.~Wellard}
\author{Lloyd~C.~L.~Hollenberg}
\affiliation{%
Centre for Quantum Computer Technology, School of Physics, The University of Melbourne, Melbourne, Victoria 3010, Australia.
}%

\date{\today}% It is always \today, today,

\begin{abstract}
Mapping the system evolution of a two-state system allows the determination of the effective system Hamiltonian directly.  We show how this can be achieved even if the system is decohering appreciably over the observation time.  A method to include various decoherence models is given and the limits of this technique are explored.  This technique is applicable both to the problem of calibrating a control Hamiltonian for quantum computing applications and for precision experiments in two-state quantum systems.  The accuracy of the results obtained with this technique are ultimately limited by the validity of the decoherence model used.

%For simple models of decoherence, this method can be applied even when the decoherence time is comparable to the oscillation period of the system. 
\end{abstract}

\pacs{03.65.Wj, 03.67.Lx}% PACS, the Physics and Astronomy

\maketitle

\section{Introduction}
Recent developments in the fabrication and control of few state quantum systems have allowed unprecedented tests of both our experimental and theoretical understanding of their behaviour.  This is typified by the continuing interest in quantum information processing and efforts to construct a quantum computer, but equally applies to other more mainstream technologies such as atomic clocks and quantum optics~\cite{Wineland:03,Meystre:99,Nielsen:00}.  

Precision control of these systems inevitably requires very accurate information about the system Hamiltonian.  Traditionally this has been obtained via a variety of experimental and theoretical techniques, though as we move toward even higher precision experiments, the need for more \emph{efficient} characterisation techniques is of utmost importance.

Recently, we introduced the concept of characterising an unknown two-state quantum system by mapping its time evolution and using this data to determine the underlying Hamiltonian~\cite{Schirmer:04,Cole:05}. Mapping the evolution in this way has the advantage that minimal knowledge of the system is required initially.  This is in contrast to tomographic methods which are in common use to map the fidelity of a given quantum operation, which assumes basic knowledge of the system, specifically the ability to rotate into different input and measurement bases~\cite{Chuang:97,James:01,Poyatos:97,Boulant:03}.   

For solid-state systems, characterisation is especially important, as the effective Hamiltonian is strongly dependant on the fabrication process and control systems.  The result is that the Hamiltonian of otherwise identical systems can vary significantly from device to device and even between different regions of the same device, making calibration or characterisation of the device a necessity.  

Typically tomography is applied to a system whose behaviour is known, but where the decoherence processes need to be characterised.  Here we take the opposite approach in which the dominant decohering mechanisms are known, but it is the strengths of the various decoherence channels that each system experiences, as well as the system control parameters which are unknown or not known to sufficient precision.  In this paper we extend our previous results~\cite{Cole:05} by demonstrating how the rate of decoherence can be characterised along with the system Hamiltonian.  

Initially, we consider the effect of a pure dephasing decoherence channel (section~\ref{sec:DecohModel}) and how this can be included in the characterisation process.  This simple model is then extended to a more general decoherence model in section~\ref{sec:moregenmodel} and we derive analytic expressions for the Fourier transform of the time evolution of the system.  Finally, in section~\ref{sec:uncerts}, we consider the way in which the resulting uncertainties scale as a function of the number of measurements and discuss the use of our analytic results in fitting experimental data.

%%%%%%%%%%%%%%%%%%%%%%%%%%%%%%%%%%%%%%%%%%%%%%%
\section{Single qubit characterisation}\label{sec:1qcharac}
Consider a two-state quantum system with a control Hamiltonian of the following form
\begin{equation}\label{eq:H}
H=\frac{d}{2}[\sin(\theta)\sigma_x+\cos(\theta)\sigma_z],
\end{equation}
using the unnormalised Pauli matrices ($\sigma$), where the $z$-axis is defined by the measurement basis. The angle $\theta$ parameterises the ratio of the $\sigma_x$ and $\sigma_z$ components and $d$ is the magnitude of the Hamiltonian.  When the control Hamiltonian is turned on, the evolution of the $z$-projection of the state of the system which is initially in the state $z(0)=1$, is given by 
\begin{equation}\label{eq:z}
z(t)=\cos(d\,t)\sin^2(\theta)+\cos^2(\theta).
\end{equation}
We have the freedom to choose the alignment of the $x$-axis and therefore do not include the effects of the $\sigma_y$ component initially by arbitrarily setting the $x$-axis to coincide with the Hamiltonian.  Once an initial axis is defined either by characterisation or on experimental grounds, then the angle between any further control Hamiltonians and the $x$-axis can be determined using additional measurements~\cite{Schirmer:04,Cole:05}.

To measure $z(t)$ experimentally, typically the system is initialised in some known state, allowed to evolve for some time $\Delta t$ and then measured.  The amount of time the system is allowed to evolve for is progressively increased in increments of $\Delta t$, giving the evolution, $z(t)$, as a function of time at $N_t$ time points over a total observation time $t_{\rm{ob}}=N_t \Delta t$.  The process must be repeated $N_e$ times to determine an ensemble average as a function of evolution time.  This is the conventional coherent oscillation experiment and results in a total of $N_T=N_e N_t$ measurements of the system which will, in general, be very large.  In this paper we will only consider so-called `strong' measurements where the measurement projects the system onto one of the two available measurement basis states.  
%%%%%%Figure showing mapping of evolution onto projection
%\begin{figure} [tb!]
%\centering{\includegraphics[width=7cm]{procdiag3.eps}} \caption
%{The $z$-projection of the time evolution of a two state system due to the effects of a control Hamiltonian is mapped using repeated cycles of initialise, evolve and measure.  After averaging over many runs, this provides the evolution of the probability distribution of the system.\label{fig:procdiag}}
%\end{figure}
%%%%%%%%%%%%%%%

\begin{figure} [tb!]
\centering{\includegraphics[width=7cm]{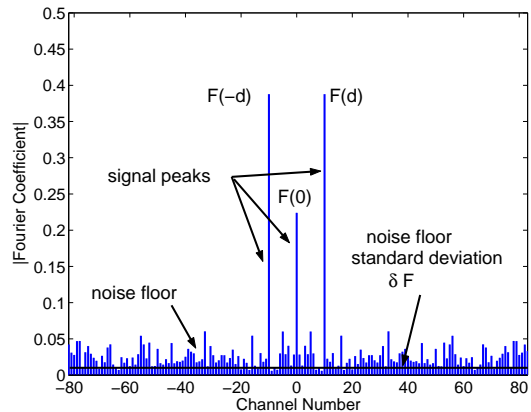}} \caption
{(Color online) Example double sided frequency spectrum showing the labelling of peaks for the case of no decoherence.  The zero-frequency peak $F(0)=F_0$ and the oscillation frequency peak $F(d)=F_p$ while the effect of only taking a finite number of measurements is to produce a noise floor.  The distribution of the noise floor gives an uncertainty estimate for the peak heights and therefore the system parameters.\label{fig:egspec}}
\end{figure}

The various parameters of the Hamiltonian can be determined from the normalised Fourier spectrum of the time evolution, $\mathcal{F}[z(t)]$.  As the evolution is purely sinusoidal, in the limit of no decoherence the Fourier spectrum is particularly simple.  It comprises $\delta$-functions at frequencies $\omega=0$ and $\omega=d$, with magnitudes $F_0=F(0)$ and $F_p=F(d)$ respectively.  An example spectrum is shown in Fig.~\ref{fig:egspec} with the appropriate features labelled.  The position of $F_p$ gives the value for $d$, whereas $\theta$ can be found from the peak heights using
\begin{equation}\label{eq:thetaF}
\cos^2(\theta)=F_0=1-2F_p.
\end{equation}

If there is an inherent measurement error probability $\eta$ (or equivalently some probability of initialisation error), this results in a reduction of the amplitude of the oscillations in a well defined manner.  This effect can be computed by calculating the sum of the Fourier spectrum, which for the case of pure delta functions is
\begin{equation}\label{eq:sumofpeaks}
\sum_\omega\mathcal{F}[z(t)]=F_0+2F_p=1-2\eta.
\end{equation}
The noise floor apparent in the Fourier spectrum stems from both the inherent noise in the experimental setup and the discretisation or `projection' noise~\cite{itano:93} due to the fact that each measurement only returns a binary result.  As the number of ensemble measurements ($N_e$) is increased, the noise floor reduces accordingly.  We can therefore use the distribution of the noise spectrum to assign uncertainty estimates to the various parameters measured from the Fourier spectrum~\cite{Cole:05}.
%%%%%%Figure showing example FFT with labels
%%%%%%%%%%%%%%%
%%%%%%Figure improvement in signal to noise with increasing measurements
%\begin{figure} [tb!]
%\centering{\includegraphics[width=7cm]{dftexample.eps}} \caption
%{The left hand plot shows an example of a sampled time signal $z(t)=[\cos(2\pi t)+1]/2$ with $N_e=1$ (a), $2$ (b), $8$ (c) and $500$ (d) measurements at each time point, where each measurement is a projection onto the (1,-1) axis.  The corresponding DFT for each signal is shown on the right.  For a given time resolution, as the number of ensemble measurements increases, the projection noise level is reduced and the signal-to-noise ratio improved.\label{fig:dftexample}}
%\end{figure}
%%%%%%%%%%%%%%%

%%%%%%%%%%%%%%%%%%%%%%%%%%%%%%%%%%%%%%%%%%%%%%%
\section{Modelling the effect of decoherence}\label{sec:DecohModel}
To model the effect of decoherence on the characterisation process, we use the Lindbladian formalism.  The time evolution of the system is then governed by the Liouville-von Neumann equation~\cite{Gardiner:91,Gardiner:04,Nielsen:00},
\begin{equation}\label{eq:mastereq}
\frac{d\rho}{dt}=-\frac{i}{\hbar}[H,\rho]+\sum_i \mathcal{L}(\rho,L_i),
\end{equation}
where $\rho$ is the density matrix of the system and 
\begin{equation}
\mathcal{L}(\rho,L_i)=L_i \rho L_i^\dag-\frac{1}{2}\{L_i^\dag L_i,\rho\},
\end{equation}
where $L_i$ is the Lindbladian operator corresponding to a particular decoherence channel.  In general a decoherence model can include several different Lindbladian operators, each of which corresponds to a different decoherence mechanism.  

While the Lindblad formalism allows the inclusion of general forms of decoherence, it still assumes the Born (weak coupling) and Markovian (uncorrelated noise) approximations.  These approximations are made when analysing decoherence for a range of systems\cite{Boulant:03,Howard:06,Barrett:03,Burkard:04}.  There are, however, some systems where it is generally believed that the Markov approximation is not valid.  A notable example of this are systems based on superconducting qubits\cite{Martinis:05,Devoret:04,Astafiev:04,Bertet:05}, where the dominant source of noise is thought to be from background charge fluctuators and/or 1/$f$ noise.  In this situation, it is often difficult or impossible to write down the evolution of the system in the form of Eq.~(\ref{eq:mastereq}) and therefore the analysis of the decoherence needs to be tailored to the particular system in question\cite{Makhlin:04,Makhlin:03,Maniscalco:06,DiVincenzo:05,Wilhelm:05}.  For simplicity, we will not treat systems of this form but merely point out that our analysis is only valid for systems whose evolution can be modelled using a master equation of Lindblad form.  Alternatively, our approach can be used to determine an effective phenomenological model for the system if Markovian evolution is assumed.

Initially we will only consider pure dephasing, as this is often considered to be the dominant form of decoherence for solid-state qubits~\cite{Makhlin:01,Khaetskii:02,Barrett:03}.  We will consider a more general model of decoherence in a later section.  The Lindbladian operator for pure dephasing is given by,
\begin{equation}\label{eq:Lsigz}
L_{z}=\sqrt{\Gamma_z}\,\sigma_z.
\end{equation}
For the case of pure dephasing, we expect an approximately exponential decay in the oscillations given by the decoherence rate, $\Gamma_z$.  As the original evolution consists of only one oscillation frequency, the exponential decay results in a broadening of the peaks in the Fourier spectrum into Lorentzian (Cauchy) distributions.  It is instructive to look at how this behaviour is modified when $\theta\neq\pi/2$, both in the time and frequency domains, for reasons which will become apparent later. 

\begin{figure} [tb!]
\centering{\includegraphics[width=8cm]{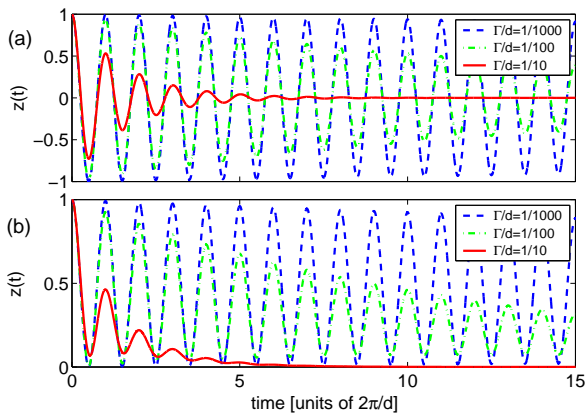}} \caption
{(Color online) The $z$-projection of the time evolution of a two state system after preparation in the $z=1$ state, for several different dephasing rates.  The magnitude of the Hamiltonian ($d=1$) is kept constant while the angle between the components is varied.  (a) For $\theta=\pi/2$, the oscillations decay exponentially.  (b) With $\theta=\pi/4$, the system undergoes a different evolution, but stills decays to the mixed state $z(\infty)=0$.  Note the difference in scales on the vertical axes in (a) and (b).\label{fig:evolexample}}
\end{figure}

Solving Eq.~(\ref{eq:mastereq}) numerically for pure dephasing, the oscillations do decay exponentially when $\theta=\pi/2$.  When $\theta\neq\pi/2$, the effect of dephasing is no longer purely exponential decay, but shows a slow decrease towards the maximally mixed state $z(\infty)=0$.   Fig.~\ref{fig:evolexample} shows the evolution of the z-projection for (a) $\theta=\pi/2$ and (b) $\theta=\pi/4$ for several different dephasing rates.  
%%%%%%Figure showing the evolution for two different values of theta
%%%%%%%%%%%%%%%
To characterise a system which is undergoing decoherence, we need to be able to account for these effects.  

\begin{figure} [tb!]
\centering{\includegraphics[width=8cm]{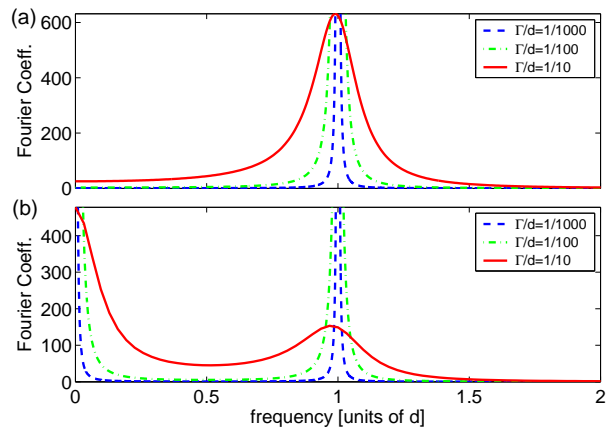}} \caption
{(Color online) The Fourier transform of the system evolution given in Fig.~\ref{fig:evolexample}, in the presence of dephasing, plotted for $\omega\ge 0$.  The magnitude of the Hamiltonian, $d$,  is kept constant and results displayed for (a) $\theta=\pi/2$, and (b) $\theta=\pi/4$.  The peak is reduced when $\theta\neq\pi/2$, as expected, and the zero-frequency component increases.  As the decoherence rate increases, the peaks broaden and the frequency shifts slightly.\label{fig:FFTexample}}
\end{figure}

Ideally we would like to solve Eq.~(\ref{eq:mastereq}) for an arbitrary Hamiltonian to obtain the time domain behaviour, though in general it is non-trivial to invert the evolution to compute the Hamiltonian parameters.  Instead, we will look at the system behaviour in the Fourier domain.  Fig.~\ref{fig:FFTexample} gives the Fourier transform of the evolution shown in Fig.~\ref{fig:evolexample} to demonstrate the dependence on both the decoherence rate and the angle $\theta$.  

Observing the peak positions under the influence of dephasing, we note that the peak position does not move appreciably.  As the peak is approximately stationary, the magnitude of the Hamiltonian vector can still be determined from its position (at least to first order).  As in the case for no decoherence, as $\theta$ is varied from $\pi/2$, the zero-frequency peak grows as expected.  The ratio of the area under the two peaks can be used to obtain a first order estimate of the value of $\theta$ while the width of the peaks are strongly dependant on the dephasing rate.
%%%%%%%%%Figure giving FFT of z(t) for theta=pi/2,pi/4 for dephasing
%%%%%%%%%%%%%%%%%%%%%%%%%%%%%%% 

In the case of negligible decoherence, the Hamiltonian parameters were determined from the Fourier transform of the system evolution and the resulting $\delta$-function peaks.  If we now consider the situation where the decoherence rate is non-negligible, but slower than the system oscillations ($\Gamma_z<d$), a broadening of the peaks in the Fourier spectrum is introduced.  
Solving Eq.~(\ref{eq:mastereq}) analytically provides a functional form which can be used to fit the experimental data.  Starting from Eq.~(\ref{eq:mastereq}), we gain insight by using the conventional variable substitution
\begin{eqnarray}
x & = & (\rho_{01}+\rho_{10})/2 \nonumber\\
y & = & (\rho_{01}-\rho_{10})/2i \label{eq:desubst}\\
z & = & \rho_{00}-\rho_{11}, \nonumber
\end{eqnarray}
to rewrite the matrix equation as a set of three coupled differential equations~\cite{Meystre:99}, where $\rho_{ij}$ are the components of the density matrix $\rho$.  For the case of pure dephasing, this gives
\begin{eqnarray} \label{eq:de_set}
\frac{dx(t)}{dt} & =&-\,d\cos (\theta )\,y(t)-2\,\Gamma_z \,x(t) \nonumber\\
\frac{dy(t)}{dt} & =&d[\sin(\theta)\, z(t)-\cos(\theta)\, x(t)] -2\,\Gamma_z \, y(t)\\
\frac{dz(t)}{dt} & =&-d\sin(\theta)\, y(t). \nonumber
\end{eqnarray}
The solution of Eqs.~(\ref{eq:de_set}) in the time domain is tractable, but does not provide any useful insight due to its complexity.  Instead we take the Fourier transform of the set of equations which results in a set of algebraic equations in terms of $X(\omega)=\mathcal{F}[x(t)]$, $Y(\omega)=\mathcal{F}[y(t)]$ and $Z(\omega)=\mathcal{F}[z(t)]$,
\begin{eqnarray}
i\omega\,X(\omega)& =&d\cos (\theta )\,Y(\omega)-2\,\Gamma_z \,X(\omega)\nonumber\\
i\omega\,Y(\omega)& =&d[\sin(\theta)\, Z(\omega)-\cos(\theta)\, X(\omega)] -2\,\Gamma_z \, Y(\omega) \label{eq:FFTeqs} \nonumber\\
i\omega\,Z(\omega) & =&-d\sin(\theta)\, Y(\omega)+C_{\mathcal{F}}.
\end{eqnarray}  
These can be solved to obtain an expression for $Z(\omega)$, where $C_{\mathcal{F}}$ is a constant of integration arising from $z(0)\neq0$.  This gives the solution to the Fourier Transform of $z(t)$ as
\begin{equation}\label{eq:Fsol}
Z(\omega)=\frac{C_{\mathcal{F}}}{i \omega+\frac{d^2 (2\,\Gamma_z+i \omega)\sin^2(\theta)}{(2\,\Gamma_z+i \omega)^2+d^2\cos^2(\theta)}},
\end{equation}
where $C_{\mathcal{F}}$ is still unknown in general.  

It is instructive to consider the case where $\theta=\pi/2$, in which we find that setting $C_{\mathcal{F}}=1$ artificially and expanding to first order around $\omega=d$, the real component of Eq.~\ref{eq:Fsol} is given by
\begin{equation}
Re[Z(\omega)]\approx\frac{\Gamma_z}{(d-\omega)^2+\Gamma_z^2},
\end{equation}
which is a Lorentzian centred about the frequency $\omega=d$ with width given by $\Gamma_z$.

Returning to the general case, if we assume the system starts in the state $x(-\infty)=y(-\infty)=z(-\infty)=0$, the input state $z(0)=1$ can be modelled as an impulse at $t=0$, resulting in the term $z(0)\delta(0)$ being added to the equation for $dz(t)/dt$.  Transforming to the frequency domain and taking into account the fact we are using the \emph{discrete} Fourier transform gives 
\begin{equation}
C_{\mathcal{F}}=\frac{z(0)}{\Delta t}=N_t \Delta \omega,
\end{equation}
as the contribution from this impulse function.

For the case of pure dephasing, the system must approach the perfectly mixed state in the long time limit, i.e.\  $x(\infty)=y(\infty)=z(\infty)=0$.  This means that the boundary condition requirements of the Fourier transform are automatically satisfied.  The steady state solution $z(\infty)=0$ also means the issues of frequency resolution and phase matching, raised in Ref.~\onlinecite{Cole:05}, are not as relevant.  As long as enough data is gathered that the steady state limit is reached, the time evolution data can be `zero-padded' to increase the frequency resolution.  If the steady state limit is not reached, then the phase difference will need to be minimised as in the case of no decoherence, though a residual error will still be present due to the amplitude mismatch.  

Providing the necessary boundary conditions are met, we now have a general procedure for dealing with a system undergoing dephasing.  We take the Fourier transform of the data, as before, and measure the peak position to determine $d$ approximately.  We then use the analytic form given in Eq.~(\ref{eq:Fsol}) and perform a nonlinear fit on $d$, $\Gamma_z$ and $\theta$ to obtain the system parameters.  To obtain a more accurate fit, the fitting process can be repeated iteratively using the same equations.  An initial estimate is obtained for each parameter and then the parameters are refitted in turn until the estimates converge.  

The parameters $d$, $\Gamma_z$ and $\theta$ predominately control the peak position, height and width respectively and the interdependence of the parameters are second order effects.  We can see this effect in the numerical results shown in section~\ref{sec:DecohModel} and it is this independence of the parameters which insures good convergence, as the covariances between the parameters are small.  The effect of imperfect measurement or initialisation can still be characterised by computing the sum of the Fourier spectrum as before, see appendix~\ref{app:etaproof}.

\section{Characterisation with a more general decoherence model}\label{sec:moregenmodel}
To treat more general decoherence, we add terms which model spontanteous absorption ($\Gamma_+$) and emission ($\Gamma_-$) (e.g.\ thermal population transfer) given by Lindbladian operators of the form
\begin{equation}
L_{\pm}=\sqrt{\Gamma_\pm}\sigma_\pm.
\end{equation}
The decoherence operator becomes
\begin{equation}
\sum_i \mathcal{L}(\rho,L_i)=\mathcal{L}(\rho,L_z)+\mathcal{L}(\rho,L_-)+\mathcal{L}(\rho,L_+),
\end{equation}
containing the three forms of decoherence which, in general, will have three different characteristic rates.  To apply the method given in section~\ref{sec:DecohModel} to the more general case requires a variable substitution to match the boundary conditions.  We define $x'(t)=x(t)-x(\infty)$, $y'(t)=y(t)-y(\infty)$ and $z'(t)=z(t)-z(\infty)$ and then solve as before.  The initial conditions must also be redefined such that $x'(0)=-x(\infty)$, $y'(0)=-y(\infty)$ and $z'(0)=z(0)-z(\infty)$.  Solving in the steady state limit, we get
\begin{eqnarray}
x(\infty) & = & \frac{2\,d\,\cos (\theta )\,y(\infty)}{4\,{{\Gamma }_z} + {{\Gamma }_+} + {{\Gamma }_-}}\\
y(\infty) & = & K \,z(\infty)\\
z(\infty) & = & \frac{\Gamma_+ -\Gamma_-}
  {{{\Gamma }_+}+{{\Gamma }_-} + d\,{\sin (\theta )}\,K}\label{eq:z_inf}
\end{eqnarray}
where
\begin{equation}
K=\frac{2\,d\,\sin (\theta )\,\left( 4\,{{\Gamma }_z} + {{\Gamma }_+} + {{\Gamma }_-} \right)}
  {4\,d^2\,{\cos^2 (\theta )} + {\left( 4\,{{\Gamma }_z} + {{\Gamma }_+} + {{\Gamma }_-} \right) }^2}.
\end{equation}
The resulting solution in the Fourier domain is
\widetext
\begin{equation}
Z'(\omega)=\frac{\left( C_{\mathcal{F}} + {{\Gamma }_+} \right) \,\left[1-z(\infty)\right] - 
    {{\Gamma }_-} \,\left[1+z(\infty)\right] 
    - d\,\sin (\theta )\,\left[L(\omega) + L^*(-\omega) \right] }{i \,\omega + \Gamma_- + \Gamma_+ + 
    \frac{2\,d^2\,M\,{\sin^2 (\theta )}}{M^2 + 4\,d^2\,{\cos^2 (\theta )}} }
\end{equation}
\endwidetext
where 
\begin{equation}
L(\omega)= \frac{\left( C_{\mathcal{F}} - i \,\omega  \right)\,\left[ y(\infty)+i \,x(\infty) \right]
               - d\,z(\infty)\,\sin (\theta )}{M - 2\,i \,d\,\cos (\theta )},
\end{equation}
\begin{equation}
M=2 \, i \, \omega+4\,\Gamma_z+ \Gamma_+ + \Gamma _-,
\end{equation}
and $L^*(\omega)$ denotes the complex conjugate of $L(\omega)$.  As $z(\infty)$ is a constant, $Z(\omega>0)=Z'(\omega>0)$ and $Z(0)=Z'(0)+z(\infty)$.  This solution is an algebraic combination of the free variables ($d$, $\theta$ and $\Gamma$'s) and can therefore be used as a fitting function for the transform of the oscillation data, as outlined previously.  

The effect of different decoherence channels can be difficult to discriminate in the Fourier domain, due to their similar action on the spectrum.  Hence a multi-parameter fit will exhibit large covariance terms, with a relatively flat potential surface of the fitting function that will not easily converge.  In this case the underlying physics of the system should be used to connect the different decoherence rates and their asymptotic values using thermodynamic, or other physical arguments.  In this way we can use the physics to provide additional constraints to improve the success of the fitting procedure,  an example of this is now given for the model in question.

\begin{figure} [tb!]
\centering{\includegraphics[width=8cm]{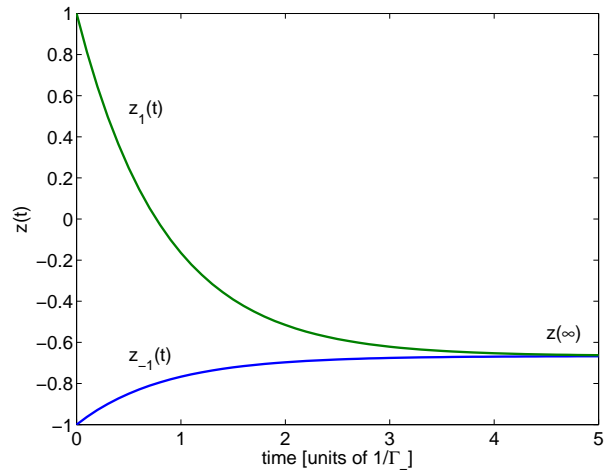}} \caption
{(Color online) The evolution of a two-state system under the influence of both spontaneous absorption and emission, in the limit of large detuning.  For this example, the emission rate is five times that of spontaneous absorption ($\Gamma_-/\Gamma_+=5$).  The path taken by the system depends on which initial state is used, though the asymptotic behaviour is the same.  The two paths are labelled $z_{-1}$ and $z_1$ depending on whether the system is initialised in the ground ($z(0)=-1$) or excited state ($z(0)=1$) respectively.\label{fig:gmgpexample}}
\end{figure}

To illustrate how this process works, we consider the special case when the measurement (initialisation) axis is coincident with the axis in which the dephasing acts.  In this situation we can measure the effect of the other (non-dephasing) decoherence terms separately and therefore reduce the number of free parameters.  By repeating the experiments as detailed earlier, in the limit of either $d=0$ or $\theta=0$,  we build up a picture of the non-Hamiltonian evolution.  As the system does not have a mechanism to move away from the $z$-axis, the influence of pure dephasing is effectively removed and the population decay results purely from the absorption and emission terms only.  This situation corresponds physically to the limit of either no driving field ($d=0$ in the rotating frame) or the large detuning limit ($\theta=0$) where the system eigenstates are coincident with the measurement basis ($\sigma_z$).   The system evolution in this limit is illustrated in Fig.~\ref{fig:gmgpexample} for an example system where $\Gamma_-/\Gamma_+=5$.  
%%%%%%%%%Figure giving FFT of z(t) for theta=pi/2,pi/4 for dephasing
%%%%%%%%%%%%%%%%%%%%%%%%%%%%%%%
Notice that the path taken by the system is different depending on which state is used for initialisation, though the steady state limit, $z(\infty)$, is the same for both.  The steady state population is given by
\begin{equation}\label{eq:zinf_gmgp}
z(\infty)=\frac{\Gamma_+-\Gamma_-}{\Gamma_++\Gamma_-},
\end{equation}
which provides one equation for determining the two rates.  Note Eq.~(\ref{eq:zinf_gmgp}) is just Eq.~(\ref{eq:z_inf}) with $d=0$ or $\theta=0$.  This means that by observing the long time behaviour \emph{only} we can reduce the number of free parameters by one.  We can use this result to define the ratio of $\Gamma_-$ to $\Gamma_+$ even in the presence of dephasing and Hamiltonian evolution.  

Observing the time behaviour of the total system, Eq.~(\ref{eq:mastereq}), with $d=0$ or $\theta=0$ provides us with another handle, as shown in Fig.~\ref{fig:gmgpexample}.  If we label the trajectories taken by the system from the two initial states as $z_{-1}$ and $z_1$ for the ground and excited states respectively, we can then fit the curves to determine both decay rates.  Alternatively, plotting the difference between the trajectories gives a simple expression,
\begin{equation}
z_1(t)-z_{-1}(t)=2\exp[-t(\Gamma_++\Gamma_-)],
\end{equation}
which can be easily fitted to determine the sum of the rates.  This can then be used with Eq.~(\ref{eq:zinf_gmgp}) to determine both rates independently without the need to fit a double exponential.  In practise, the experiment could be conducted by binning each measurement result based on the previous measurement and therefore the initialisation state.  This would save time in the initialisation phase as single qubit operations could then be minimised.

Using this type of auxiliary experiment, the number of free parameters in the expression for $Z(\omega)$ can be reduced, resulting in better convergence during the fitting process.  This in turn gives higher precision estimates for the system parameters for the same number of measurements.

\section{Estimating the uncertainty}\label{sec:uncerts}
\begin{figure} [tb!]
\centering{\includegraphics[width=8cm]{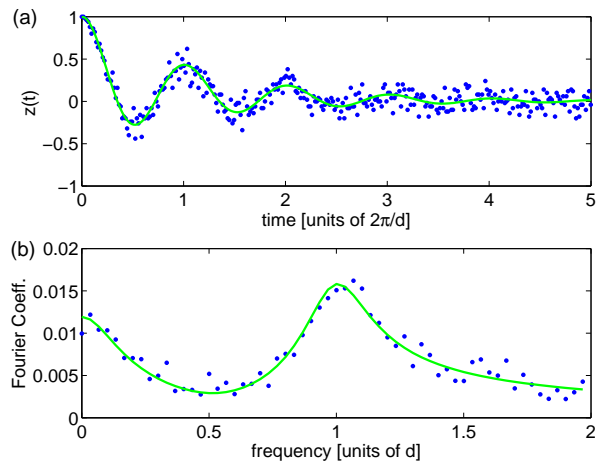}} \caption
{(Color online) Simulated data for evolution due to the example Hamiltonian, $H=0.93 \sigma_x+0.38 \sigma_z$.  (a) The time series data for $N_e=50$ (points) which approximates the ensemble average of the evolution (solid line). (b) The Fourier transform of the time domain data (points) plotted with the fitted function (solid line) using the estimated parameters in Table~\ref{tb:examplevalues}.\label{fig:examplefit}}
\end{figure}

To illustrate the method developed so far, we simulate an experiment with an arbitrary example Hamiltonian $H=0.93 \sigma_x+0.38 \sigma_z$ ($d=1$, $\theta=1$) and include a pure dephasing term with decoherence rate $\Gamma/d=0.1$.  We numerically solve the system evolution for $t_{\rm{ob}}=15$ and then include the effect of finite measurement by simulating projective measurement, using $N_t=1000$ and $N_e=50$.  The resulting Fourier spectrum is then fitted using the Levenberg-Marquardt nonlinear regression algorithm~\cite{Marquardt:63, Press:96} to perform a non-linear fit in the Fourier domain, see Fig.~\ref{fig:examplefit}.  

Using conventional nonlinear fitting routines to fit the measured data with the functions developed in section~\ref{sec:moregenmodel} has the advantage that these routines also provide uncertainty estimates based on the goodness-of-fit.  These uncertainties can be used to directly estimate the uncertainty in the final Hamiltonian parameters using the same relations derived for no decoherence\cite{Cole:05}.  We obtain rough estimates for the various parameters and then perform an iterative fitting process where each parameter is varied in turn.  Generally, convergence is achieved within 2-3 interations, although this depends on the number of measurements and ultimately on the quality of the experimental data.  The numerical values obtained from the data shown in Fig.~\ref{fig:examplefit} are given in Table~\ref{tb:examplevalues}, for an example run.  The true value ($x$), the estimate ($\hat{x}$) and the $3\sigma$ confidence interval ($\delta x$) are given for $d$, $\theta$ and $\Gamma_z$.

%%%%%%example plot
%%%%%%%
%%%%%%%
\begin{table}[tb!]
        \begin{tabular}{|c|c|c|c|c|}
        \hline
         & $x$ & $\hat{x}$ & $\delta x$ & $\delta x/\hat{x}$ \\
        \hline
        $d$ & $1.000$ & $0.996$ & $0.020$ & $0.020$\\
        $\theta$ & $1.000$ & $1.007$ & $0.030$ & $0.030$ \\
        $\Gamma_z$ & $0.100$ & $0.102$ & $0.010$ & $0.098$ \\
        \hline
        \end{tabular}
\caption{Example values from a simulated run of the fitting procedure discussed above, using the data shown in Fig.~\ref{fig:examplefit}.  The true value  ($x$), its estimate ($\hat{x}$), the uncertainty ($\delta x$) and the fractional uncertainty ($\delta x/\hat{x}$) are given for the three system parameters $d$, $\theta$ and $\Gamma_z$ with $N_t=1000$ and $N_e=50$.\label{tb:examplevalues}}
\end{table}
%%%%%%

\begin{figure} [tb!]
\centering{\includegraphics[width=8cm]{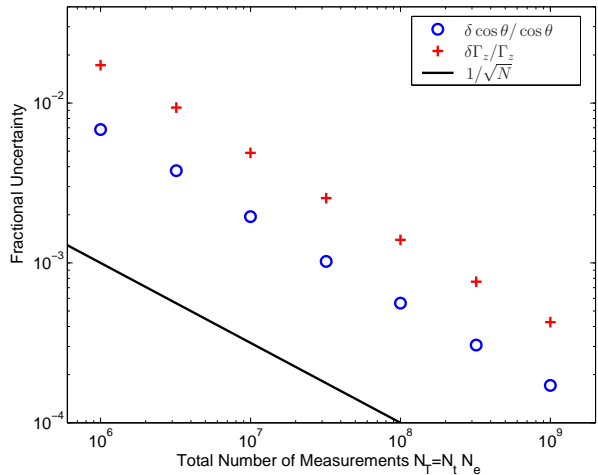}} \caption
{(Color online) The uncertainty estimate as a function of number measurements for an example Hamiltonian $H=0.93 \sigma_x+0.38 \sigma_z$ undergoing pure dephasing in the bare qubit basis.  The fractional uncertainty for the $1\sigma$ confidence interval is plotted for both the $\sigma_z$ component of the Hamiltonian and the dephasing rate $\Gamma_z$.  The scaling is approximately proportional to $1/\sqrt{N_T}$ and the absolute fractional uncertainty is independent of the decoherence rate.\label{fig:uncert_scaling}}
\end{figure}

As in the case of no decoherence, we are ultimately interested in how the parameter uncertainties scale with increasing number of measurements.  Fig.~\ref{fig:uncert_scaling} shows the fractional uncertainty ($1\sigma$ level) for the $\sigma_z$ component of the Hamiltonian and the decoherence rate of the example system, as a function of the number of measurements.  Similar to the case for no decoherence, the frequency uncertainty is predominantly controlled by the time resolution and is typically much smaller than the other uncertainties.  Using a constant number of time points ($N_t=1\times10^6$) and increasing the number of ensemble measurements ($N_e$) we find that the fractional uncertainty scales proportional to $1/\sqrt{N_T}$.  The fractional uncertainty is approximately a factor of $7$ larger than for no decoherence, which is equal to $1/\sqrt{N_T}$~\cite{Cole:05}.  This represents the penalty for fitting including the decoherence terms.  The uncertainty does not change appreciably when the decoherence rate changes, with the curves for $\Gamma/d=0.1$, $\Gamma/d=0.01$ and $\Gamma/d=0.001$ giving identical behaviour.
%fit and scaling plots
%%%%%%%%%figure showing scaling of uncertainty with number of measurements
%%%%%%%%%%%%%%%%%%%%%%%%%%%%%%%

The limiting factor in both the fitting procedure and the uncertainty analysis is, `how appropriate is the decoherence model?'  If the model used is not suitable, this will be apparent as the fitting procedure will not satisfactorily converge, even after many iterations.  This can be most easily determined by inspection of the Fourier spectrum shape, when compared to that of the closest fitting parameters.  If the model is found to be the limiting factor in the parameter estimation, then a more sophisticated model is required which may require additional experiments.  An example of this type of additional experiment is the process discussed earlier for determining spontaneous absorption and emission terms independently  from the effects of dephasing.  In situations where the Born and/or Markov approximations break-down, it may not be possible to model the decoherences using a closed-form expression for the time evolution, such as Eq.~(\ref{eq:mastereq}).  In this case, a more sophisticated analysis would be required to determine the exact decoherence processes.

\section{Conclusion}
Mapping the time domain evolution of a two state system has previously been shown to provide a systematic method for characterising the system Hamiltonian for quantum information processing applications.  The major drawback being the long decoherence times required to accurately map the system evolution.  In this paper, we have shown how this technique can be extended to include the case where the coherent oscillations are damped due to the effects of decoherence.  

The effects of relaxation and dephasing are considered and a procedure developed to incorporate general decoherence models.  This provides an `after the fact' analysis technique which also allows for arbitrary accuracy characterisation given enough measurement data.  The technique can be applied to two-state experiments other than those required for quantum computing applications and the achievable precision is only limited by the validity of the decoherence model used.

The procedure requires deriving a fitting function based on a model for the decoherence processes affecting the system.  We derive the result for a simple model including dephasing and spontaneous absorption and emission, acting in the measurement basis of the physical qubit.  Other more complex models can be included, as long as the model can be written down as a closed form master equation.  In cases whether the decoherence cannot be written in this form, the procedure can be used with an effective model and the convergence of the parameter estimates provides an indication of the validity of this effective model.

The complexity of the decoherence model which can be employed is restricted by the need to fit the resulting measurement data.  If too many degrees of freedom are introduced into the model, the fitting procedure will not converge sufficiently to provide usable parameter estimates.  In this situation, we find that using the physics of the system and performing other auxiliary experiments can provide additional constraints on the parameters and therefore reduce the uncertainties.  This technique is ultimately limited by the accuracy of the decoherence model, the stability of the Hamiltonian and the number of measurements taken.

\begin{acknowledgments}
JHC and LCLH would like to acknowledge helpful discussions with F. Wilhelm, thank the von Delft group at LMU for their hospitality, and acknowledge financial support from the DFG through the SFB631.  LCLH was supported by the Alexander von Humboldt Foundation.  JHC, SGS and DKLO acknowledge funding from the Cambridge-MIT institute, Fujitsu and the EPSRC IRC QIP.  DKLO also thanks Sidney Sussex College for support and acknowledges EU grants TOPQIP (IST-2001-39215) and RESQ (IST-2001-37559).
This work was supported in part by the Australian Research Council, the Australian government, the US National Security Agency, the Advanced Research and Development Activity and the US Army Research Office under contract number W911NF-04-1-0290.  
\end{acknowledgments}

\appendix
\section{Determination of an initialisation or measurement error}\label{app:etaproof}
First we show that the initial state of the system, $z(0)$, can be determined from the Fourier transform of the data.  The definition of the discrete Fourier transform is given by
\begin{equation}
\mathcal{F}[z(k)]=\sum_{k=0}^{N-1}z(k\,\Delta t)e^{i\omega k n/N},
\end{equation}
where $N$ is the number of time or frequency channels, $k$ denotes the channel number of the time series, $n$ denotes the frequency channel number and both $n$ and $k$ are integers.
If we compute the sum of the discrete Fourier spectrum over all frequency channels, this gives
\begin{equation}
\sum_{n=0}^{N-1}\mathcal{F}[z(t)]=\sum_{n=0}^{N-1}\sum_{k=0}^{N-1}z(k\,\Delta t)e^{i\omega k n/N}.
\end{equation}
Interchanging the order of the summations gives
\begin{equation}
\sum_{n=0}^{N-1}\mathcal{F}[z(t)]=\sum_{k=0}^{N-1}z(k\,\Delta t)\sum_{n=0}^{N-1}e^{i\omega k n/N},
\end{equation}
as $z(k\,\Delta t)$ is independent of $n$.
If we consider the inner summation term, evaluating the real and imaginary parts separately gives
\begin{equation}
\sum_{n=0}^{N-1}e^{i\omega k n/N}=\sum_{n=0}^{N-1}\cos(\omega k n/N)+i\sum_{n=0}^{N-1}\sin(\omega k n/N).
\end{equation}
When $k=0$, the cosine term is
\begin{equation}
\left.\sum_{n=0}^{N-1}\cos(\omega k n/N)\right| _{k=0}=N,
\end{equation}
whereas, when $k>0$ the summation over an entire period cancels out for suitable values of $N$, giving
\begin{equation}
\left.\sum_{n=0}^{N-1}\cos(\omega k n/N)\right| _{k>0}=0.
\end{equation}
Similar, the imaginary sine term is equal to zero for all values of $k$.
Putting this together we find that the inner summation is equal to 
\begin{equation}
\sum_{n=0}^{N-1}e^{i\omega k n/N}=N\delta_{k,0}
\end{equation}
i.e.\ the Kronecker delta function.
This then gives us
\begin{equation}
\sum_{n=0}^{N-1}\mathcal{F}[z(k\,\Delta t)]=N\sum_{k=0}^{N-1}z(k\,\Delta t)\delta_{k,0}=N z(0),
\end{equation}
so the initial state of the system is given by the sum over the discrete Fourier transform of the evolution, divided by the number of time points.  If the system is under going decoherence, the result is the same as long as the boundary conditions are still satisfied.  If $z(\infty)\neq0$ then the variable substitution discussed in section~\ref{sec:moregenmodel} can be used and the sum computed over the Fourier transform of $z'(t)$.

If we consider some probability $\eta$ that the system is initialised in the incorrect state (modelled by a `bit-flip' error immediately after initialisation), $\langle\psi(0)|\psi(0)\rangle=(1-\eta)\langle0|0\rangle+\eta\langle1|1\rangle$, then the initial state becomes $z(0)=1-2\eta$.  Therefore the sum over all frequencies,
\begin{equation}
\sum_{n=0}^{N-1}\mathcal{F}[z(k\,\Delta t)]=1-2\eta,
\end{equation}
gives the initialisation error.

For some system evolution given by $U=\exp(-i Ht)$, where $H$ is given by Eq.~(\ref{eq:H}), the evolution of the $z$-projection is given by
\begin{equation}
z(t)=\rm{Tr}[\sigma_z U \rho(0) U^\dag],
\end{equation}
where $\rho(0)=|0\rangle\langle0|$ is the starting state, $z(0)=1$.  
If we model an initialisation error as a bit flip before the evolution, this gives the evolution of the `system+error' as 
\begin{equation}
z_{\rm{err}}^{\rm{pre}}(t) = \rm{Tr}[\sigma_z U \sigma_x^\dag \rho(0) \sigma_x U^\dag]=-z(t).
\end{equation}
Similarly we find that applying the bit flip error after the evolution gives
\begin{equation}
z_{\rm{err}}^{\rm{post}}(t) = \rm{Tr}[\sigma_z \sigma_x^\dag U \rho(0) U^\dag \sigma_x]=-z(t).
\end{equation}
The two error locations are therefore equivalent for the Hamiltonian given in Eq.~(\ref{eq:H}) and the resulting system evolution with an error probability $\eta$ is given by
\begin{equation}
z_\eta(t)=(1-\eta)z(t)+\eta z_{\rm{err}}(t)=(1-2\eta)z(t),
\end{equation}
which is the measurement error model used previously~\cite{Cole:05}.  This means that the sum over the Fourier transform gives the probability of a measurement or initialisation error or the cumulative effect if both are present.

%\newpage %Just because of unusual number of tables stacked at end
\bibliography{hcwd}% Produces the bibliography via BibTeX.

\end{document}